\documentclass[10pt,aps,prd,preprint,onecolumn]{revtex4}

\usepackage{amssymb}
\usepackage{amsmath}
\usepackage{graphics}
\usepackage{latexsym}
\usepackage{array}
\usepackage[activeacute,english]{babel}
\usepackage[latin1]{inputenc}
\usepackage[dvips]{graphicx}
\usepackage{color}
\usepackage{epstopdf}
\usepackage{amssymb}
\usepackage{slashed}
\usepackage{mathrsfs}
\usepackage{amsmath}
\usepackage{wasysym}

\newcommand{\bn}{\begin{eqnarray}}
\newcommand{\en}{\end{eqnarray}}
\newcommand{\be}{\begin{equation}}
\newcommand{\ee}{\end{equation}}

\newcommand{\bc}{\begin{center}}
\newcommand{\ec}{\end{center}}

\begin{document}

\title{\Large \bf Neutrino dispersion relations at finite temperature and density in
the Left-Right Symmetric Model}

\author{F. D. Villalba-Pardo}
\email{fdvillalbap@unal.edu.co}

\affiliation{Departamento de F\'{\i}sica, Universidad Nacional de Colombia.\\
Ciudad Universitaria, Bogot\'{a} D.C., Colombia.}

\author{C. J. Quimbay\footnote{Associate researcher of Centro
Internacional de F\'{\i}sica, Bogot\'{a} D.C., Colombia.}}
\email{cjquimbayh@unal.edu.co}

\affiliation{Departamento de F\'{\i}sica, Universidad Nacional de Colombia.\\
Ciudad Universitaria, Bogot\'{a} D.C., Colombia.}

\date{\today}

\begin{abstract}
In this work we calculate the most general left-handed neutrino
thermal self-energy at one-loop order in perturbation theory using
the Mellin summation technique. We perform this calculation in the
real-time formalism of quantum field theory at finite temperature
and density assuming that there exists an excess of leptons over
antileptons in the medium. Thus, we obtain a novel general
expression for the left-handed neutrino effective thermal mass
which depends on lepton masses, boson masses, leptonic chemical
potential and temperature. As an application of these results into
the context of the Left-Right Symmetric Model, we calculate the
left-handed neutrino dispersion relations and we obtain the
corresponding effective thermal masses for the unbroken,
parity-broken and fully-broken symmetry phases.

\end{abstract}


\maketitle


\section{Introduction\label{sec:intro}}

\par The observational evidence obtained from neutrino oscillation
experiments is well understood in terms of massive and mixed
neutrinos \cite{Beringer2012}. The search of a satisfactory
explanation about the origin of neutrino masses and mixings is one
of the main interests of neutrino physics current research. Since
left-handed neutrinos are massless in the Electroweak Standard
Model (ESM), then neutrino oscillations are the first
phenomenological evidence of physics beyond the ESM. In models
beyond the ESM that include a neutrino flavor mixing matrix, the
explanation of the small masses of left-handed neutrinos is
possible after the implementation of the see-saw mechanism.

\par The fact that neutrinos are massive and mixed has consequences
in different scenarios in which these particles participate. One
of those scenarios is the early universe in which neutrinos can
influence the physical processes involved in primordial
nucleosynthesis, leptogenesis and the formation of the large scale
structure of the universe \cite{Mohapatra2004}. In the
astrophysical context, we know that neutrinos can influence
strongly the dynamics of supernovae and neutron stars. One of the
main aspects that must be taken into account in the analysis of
this kind of processes is that neutrinos are propagating through a
medium which interacts with them. As shown in the literature
\cite{Arteaga2007,Weldon1982}, the medium changes the properties
of propagation of neutrinos. The collective effects of the medium
in neutrinos can be described by means of thermal effective masses
which are obtained from the dispersion relations. Given that
neutrino oscillation patterns in a medium are a function of
neutrino masses, then thermal and density effects of the medium in
neutrino oscillations are described by the neutrino thermal
effective masses \cite{Stodolsky1987}-\cite{McKellar1993}.

\par The main goal of this work is to calculate the most general
left-handed neutrino thermal self-energy at one-loop order in
perturbation theory. This calculation is performed in the
real-time formalism of quantum field theory at finite temperature
and density considering that neutrinos are propagating in a
medium. The medium is constituted by massive leptons, massive
scalar bosons and massive gauge bosons. This medium is
characterized by leptonic chemical potentials which are associated
to the excess of leptons over antileptons in the medium. From this
calculation, we obtain a novel general expression for the
left-handed neutrino effective thermal mass which depends on
lepton masses, boson masses, leptonic chemical potential and
temperature. To illustrate the applicability of these results, we
calculate the left-handed neutrino dispersion relation at finite
temperature and density in the framework of the left-right
symmetric model (LRSM). Starting from this neutrino dispersion
relation, we obtain the left-handed neutrino effective masses at
finite temperature and density for the unbroken, parity-broken and
fully-broken phases.

\par A first work about the calculation of fermionic dispersion
relations at finite temperature was presented in
\cite{Weldon1982a}. In that work, for a non-abelian gauge theory,
the dispersion relations of massless fermions were calculated at
one-loop order in perturbation theory using the real time
formalism of the quantum field theory at finite temperature.
Additionally, a description about possible modifications of the
fermionic dispersion relations for the case of gauge theories with
parity and chirality violation was also presented in that work.
Later in \cite{Quimbay1995}, the fermionic dispersion relations at
finite temperature were calculated in the context of the ESM for a
plasma with vanishing fermionic chemical potentials. Particularly
in \cite{Quimbay1995}, the left-handed neutrino dispersion
relations were calculated using an analytical approach for the
case where particles are massless (electroweak unbroken phase),
while for the case where charged leptons, electroweak gauge bosons
and the Higgs boson are massive (electroweak broken phase) these
dispersion relations were calculated using a numerical approach
due to the difficulty to develop an analytical calculation.
Posteriorly in \cite{Quimbay1999}, the left-handed neutrino
dispersion relations at finite temperature and density were
analytically calculated in the electroweak unbroken phase of the
ESM for a plasma characterized by non-vanishing leptonic chemical
potentials. In this last reference, the left-handed neutrino
effective masses were obtained as functions of the temperature and
the leptonic chemical potentials.

\par In contrast with the calculations performed previously in
\cite{Quimbay1995,Quimbay1999}, in this work we calculate
analytically the most general left-handed neutrino thermal
self-energy for the case of a medium constituted by leptons and
bosons with mass and where the medium is characterized by
non-vanishing leptonic chemical potentials. We show that the
left-handed neutrino effective thermal mass obtained from this
most general self-energy can be reduced for specific cases to the
form of effective thermal masses known in the literature.

\par Different aspects related to the dispersion relations
of neutrinos propagating in a medium have been studied previously.
For instance, some calculations in the context of the ESM have
been performed in order to obtain the refraction index
\cite{Nieves1989}, the effective potential \cite{DOlivo1992}, the
damping rate \cite{Tututi2002} and some properties of the
collective excitations of neutrinos in a medium
\cite{Boyanovsky2005}. The study of fermionic dispersion relations
at finite temperature and density in other frameworks different to
the ESM has been also considered. Particularly, in the context of
a minimal supersymmetric extension of the ESM, it have been
studied some electromagnetic properties of neutrinos in a medium
\cite{Masood1993}. Additionally in \cite{Wizansky2006} the finite
temperature corrections to the density of neutralinos, which could
be relevant to the dark matter problem, were calculated. Another
relevant work was presented in \cite{Riotto1997}, where the
propagation of Majorana fermions at finite temperature and density
was studied and the results were applied to describe the
thermodynamical properties of a system with neutralinos in the
framework of a minimal supersymmetric extension of the ESM.

\par This paper is organized as follows. Firstly, in section
\ref{sec:lrsm}, we show some features of the LRSM which are
relevant to our analysis. Next, in section
\ref{sec:dispersionrelations}, the most general left-handed
neutrino thermal self-energy is analytically calculated using the
Mellin summation technique for the case in which the thermal
medium is constituted by massive leptons, massive gauge bosons,
massive scalar bosons, and assuming that there are a leptonic
chemical potential associated to an excess of leptons over
antileptons in the medium. Furthermore, in section
\ref{sec:effectivemasses}, a general expression for the
left-handed neutrino effective thermal mass is obtained and as an
application of this result, we obtain the left-handed neutrino
effective thermal masses for the unbroken, parity-broken and
fully-broken symmetry phases in the context of the LRSM. Finally,
in section \ref{sec:conclusions}, we present some conclusions.


\section{Relevant aspects of the LRSM\label{sec:lrsm}}

\par The LRSM is an extension of the ESM which
is based on the gauge symmetry group $SU(2)_L\times SU(2)_R\times
U(1)_{B-L}$ and whose origin was motivated by the fact that parity
should be restored at high energies \cite{Mohapatra2004}. This
extension is supported by the existence of a right-handed gauge
group $SU(2)_R$ that has the same running coupling constant of the
left-handed gauge group $SU(2)_L$. The $SU(2)_R$ gauge group
introduces three right gauge bosons which are coupled to the
right-handed fermions of the model. By this reason, in the LRSM,
it is assumed that there exist right-handed neutrinos and then
left-handed neutrinos are massive as a consequence of the see-saw
mechanism. The gauge coupling constants of this model $g$ and
$g^\prime$ are associated to the gauge groups $SU(2)_{R,L}$ and
$U(1)_{B-L}$ respectively. The parametrization of these gauge
coupling constants in terms of the electroweak angle $\theta_w$
and the electric fundamental charge $e$ is given by
$\sin\theta_w=g^\prime/\sqrt{g^2+2{g^\prime}^2}$ and $e=g
g^\prime/\sqrt{g^2+2{g^\prime}^2}$. The interaction Lagrangian
density for the leptonic sector written in terms of the physical
gauge bosons is
\begin{eqnarray}
{\cal L}=&&-\frac{g}{\sqrt{2}}\left(\overline{\nu_{lL}}\gamma^\mu
l_LW^+_ {L\mu}+\overline{l_L}\gamma^\mu\nu_{lL}W^-_{L\mu}\right)+
\overline{\nu_{lL}}\gamma^\mu\nu_{lL}\left[\frac
g2(c_w+s_wt_w)Z_\mu+
\frac{g'}{2}t_wZ'_\mu\right]\nonumber\\
&&-\frac{g}{\sqrt{2}}(\overline{N_{lR}}\gamma^\mu l_R W^+_{R\mu}+
\overline{l_R}\gamma^\mu
N_{lR}W^-_{R\mu})+\overline{N_{lR}}\gamma^\mu N_{lR}
\frac12\left(g\frac{\sqrt{c_{2w}}}{c_w}+g't_w\right)Z'_\mu,\label{eq:gaugeleftright}
\end{eqnarray}
where $l$ runs over the three lepton flavors $e, \mu, \tau$. In
this Lagrangian density, $\nu_{lL}$ represents the left-handed
neutrino fields, $N_{lR}$ the right-handed neutrino fields, $l_L$
the left-handed charged lepton fields, $l_R$ the right-handed
charged lepton fields, $W^{\pm}_{L\mu}$ the charged left gauge
boson fields, $W^{\pm}_{R\mu}$ the charged right gauge boson
fields, $Z$ the neutral left gauge boson field, $Z'$ the neutral
right gauge boson field, $c_w=\cos\theta_w$ and
$c_{2w}=\cos(2\theta_w)$. Scalar multiplets are introduced in the
LRSM in such a way that parity is broken at a very high energy
scale when the $SU(2)_R$ symmetry is spontaneously broken. The
spontaneous breaking of the $SU(2)_L$ symmetry occurs at the
electroweak scale and the ESM phenomenology is recovered. The
simplest way to obtain a consistent pattern of symmetry breaking,
allowing the existence of right-handed Majorana neutrinos, is to
include the following scalar multiplets \cite{RodriguezY2002}: A
bidoublet $\Phi \equiv (2,2,0)$ and two triplets $\Delta_R \equiv
(1,3,2)$, $\Delta_L \equiv (3,1,2)$. The Yukawa Lagrangian density
for leptons is written in terms of these scalar multiplets as
follows
\begin{eqnarray}
-{\cal L}_Y=&&\sum_{i,j}\left[\tilde
h_{ij}\overline{\Psi_{Li}}\Phi\Psi_{Rj}+
\tilde g_{ij}\overline{\Psi_{Li}}\tilde\Phi\Psi_{Rj}\right.\nonumber\\
&&\left.+f_{ij}(\overline{\Psi^c_{Ri}}\vec\tau\cdot\vec\Delta_L\Psi_{Lj}+
\overline{\Psi^c_{Li}}\vec\tau\cdot\vec\Delta_R\Psi_{Rj})\right]+h.c,\label{eq:yukawaleftright}
\end{eqnarray}
where the sum runs over the three lepton flavors, $\Psi_{L}^T =
(\nu_L, l_L)$ is the left-handed lepton doublet and $\Psi_{R}^T =
(N_R, l_R)$ is the right-handed lepton doublet. After spontaneous
symmetry breaking the scalar fields gain a non-vanishing vacuum
expectation value \cite{RodriguezY2002}
\begin{equation}
\langle\Phi\rangle=\frac{1}{\sqrt{2}}\left(
\begin{array}{cc}
k_1e^{i\alpha}&0\\
0&k_2
\end{array}
\right),
\end{equation}
\begin{equation}
\langle\Delta_L\rangle= \frac{1}{\sqrt{2}}\left(
\begin{array}{cc}
0&0\\v_L&0
\end{array}
\right),
\end{equation}
\begin{equation}
\langle\Delta_R\rangle= \frac{1}{\sqrt{2}}\left(
\begin{array}{cc}
0&0\\v_Re^{i\theta}&0
\end{array}
\right),
\end{equation}
where $k_1$, $k_2$, $v_L$, $v_R$, $\alpha$ and $\theta$ are real
numbers. The values of the vacuum expectation values $k_1$, $k_2$,
$v_L$ and $v_R$ are restricted by some constraints. For instance,
$k_1$ and $k_2$ are related as $k_1^2 + k_2^2 \simeq (246$
GeV$)^2$. The experimental condition $M_{W_L}^2/M_{Z_L}^2 \simeq
\cos^2(\theta_w)$ implies that $v_L$ should be much smaller than
$k_1$ and $k_2$ \cite{Deshpande1991,Baremboim1998}. Additionally,
$v_R$ must be at least of the order of $10^7$ GeV to give heavy
masses to the right-weak bosons $W_R^+, W_R^-, Z_R^0$. In this
way, $v_R$ is consistent with the lowest experimental limits for
these masses \cite{RodriguezY2002}. Given the present experimental
bounds on neutrino masses \cite{Beringer2012} ($\sum_\nu m_\nu<1$
eV) and supposing the Majorana-Yukawa couplings are of order
unity, we have that $v_R$ must be of the order of $10^{15}$ GeV to
obtain the mass of the right-handed heavy neutrinos at this order.
The genuine spontaneous CP-violation phases $\alpha$ and $\theta$
are obtained after we take similarity transformations over the
scalar fields and some of the four phases involved in the scalar
fields are absorbed by defining specific forms of the
transformation matrices \cite{RodriguezY2002}. As a consequence
right-handed and left-handed neutrinos gain mass. The resulting
mass matrix can be block-diagonalized and one of such blocks is
the mass matrix for right-handed heavy neutrinos $M^{heavy}\approx
f_{diag}v_R$, with $f_{diag}$ representing the diagonalized form
of the Majorana-Yukawa coupling matrix $f_{ij}$. The other block
corresponds to the mass matrix for the left-handed light neutrinos
and can be written, in the limit $v_R\gg k_1, k_2$, as
\cite{RodriguezY2002}
\begin{equation}
M_\nu\approx f v_L +
(k_1e^{i\alpha}h+k_2g)(f_{diag}v_Re^{i\theta})^{-1}(k_1e^{i\alpha}h^t+k_2g^t).
\end{equation}
Then a see-saw pattern of masses for neutrinos is obtained in this
model. It is important to note that using other successful
neutrino mass generation mechanisms in the LRSM, as for instance
the dynamical left-right symmetry breaking \cite{Akhmedov:1995vm},
it is possible to obtain that the right-handed scale in this model
could be of the order of 20 $TeV$.


\section{Dispersion relations for left-handed neutrinos\label{sec:dispersionrelations}}

\begin{figure}[H]
\includegraphics[width=0.45\textwidth]{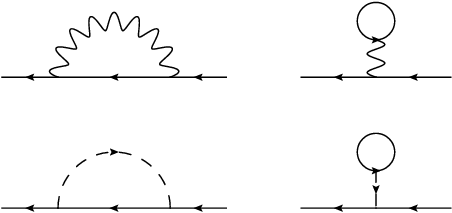}
\caption{General diagrams for the computation of the fermion
    self-energy.}
\label{figura1}
\end{figure}

As it has been discussed in the literature \cite{Arteaga2007}, the
emergence of thermal effective masses for particles in a medium
can be seen as a consequence of the existence of a privileged
frame of reference for the system, which is called the rest frame
of the plasma. This lack of invariance makes that the poles of the
propagators of fermions and bosons can no longer be located at the
physical mass but at a shifted value. Therefore, we must consider
the modifications of the fermion propagator poles due to the
medium at finite temperature and density in order to find the
effective masses of neutrinos. These effective masses can be
obtained from the calculation of the neutrino thermal self-energy
at one-loop order in perturbation theory. To compute this
self-energy, we use the real-time formalism of the quantum field
theory at finite temperature and density. Using this formalism, it
is possible to separate directly the quantum and thermal
contributions to the fermion self-energy. The one-loop order
Feynman diagrams used for this calculation have the generic form
shown in figure \ref{figura1}.

The Feynman rules in the real-time formalism are essentially the
same as those in the imaginary-time formalism, the differences are
mainly in the form of the propagators. The propagator for fermions
of mass $M_F$ is written as
\begin{equation}\label{eq:fermionprop}
S(q)=({\not
q}+M_F)\left[\frac{i}{q^2-M_F^2+i\epsilon}+\Gamma_f(q)\right],
\end{equation}
whereas the propagators for gauge bosons (in a generalized
$\xi$-gauge) and scalars of mass $M_B$ are
\begin{eqnarray}
D_{\mu\nu}(p)=&&\left(-g_{\mu\nu}+\frac{p_\mu
p_\nu(1-\xi)}{p^2-\xi M_B^2}
\right)\left[\frac{i}{p^2-M_B^2+i\epsilon}-\Gamma_b(p)\right],\label{eq:vectorprop}\\
D(p)=&&\frac{i}{p^2-M_B^2+i\epsilon}-\Gamma_b(p),\label{eq:scalarprop}
\end{eqnarray}
respectively. The functions $\Gamma_{b,f}$ depend on temperature
$T$ and leptonic chemical potential $\mu$. These functions are
given by
\begin{equation}\label{eq:gammadef}
\Gamma_{b,f}(p)=2\pi\delta(p^2-M_{B,F}^2)n_{b,f}(p),
\end{equation}
where $n_{b}$ represents the Bose-Einstein distribution and
$n_{f}$ the Fermi-Dirac distribution. These distributions are
given by
\begin{eqnarray}\label{eq:distributions}
n_b(p)=&&\frac1{\exp\left(\frac{p^0}{T}\right)-1},\\
n_{f}(p)=&&\theta(p\cdot u)n_{f}^-(p)+\theta(-p\cdot u)n_{f}^+(p),\\
n_f^\pm(p)=&&\frac1{\exp\left(\frac{p^0\pm\mu}{T}\right)+1}.
\end{eqnarray}

We perform the calculation in the Feynman gauge, {\it i. e.} we
fix $\xi=1$ into (\ref{eq:vectorprop}). However, as we will
discuss later, the validity of our results is independent of the
gauge fixing. As it was discussed in \cite{Weldon1982}, it is
enough to consider the real part of the finite temperature
contribution to the fermion self-energy to be able to calculate
the fermionic dispersion relation and to obtain the thermal
effective mass. At this point, it is necessary to specify the
phase of symmetry breaking of the LRSM in which we are working.
For the sake of generalization, we will first consider the case in
which left-right symmetry is fully broken in such a way that all
the fermions and bosons of this model gain mass. Given the chiral
nature of the interactions in this case, we expect that the real
part of the self-energy for fermions of momentum $K_\mu = (\omega,
{\vec k})$ propagating through a medium with four-velocity $u_\mu
=(1,{\vec 0})$ can be parametrized by virtue of the Lorentz
covariance as \cite{Weldon1982a}
\begin{equation}\label{eq:anotherparamself}
Re\Sigma'=-{\not K}(a_{L}L+a_{R}R)-{\not u}(b_{L}L+b_{R}R)-
M_F(c_{L}L+c_{R}R),
\end{equation}
where $L \equiv (1-\gamma^5)/2$ and $R \equiv (1+\gamma^5)/2$ are
the left- and right-handed chiral projectors respectively, and
$a_{L,R}$, $b_{L,R}$ and $c_{L,R}$ are the Lorentz invariant
functions associated to left-handed (subindex $L$) and
right-handed (subindex $R$) fermions. These invariant functions
depend on the scalars $K^2=\omega^2 - k^2$ and $K\cdot u=\omega$,
and they are associated to contributions coming from interactions
acting over fermions of each chirality. In \cite{Quimbay1995}, it
was shown that the presence of a term proportional to the fermion
mass into (\ref{eq:anotherparamself}) makes impossible to
establish a analytical dispersion relation for fermions of each
chirality. However, as will be discussed later, the smallness of
the left-handed neutrino masses causes the deviation from the
chiral behavior of the dispersion relations to be very small.
Therefore, we can neglect safely the third term in
(\ref{eq:anotherparamself}) by assuming that $M_F = 0$ and then
the real part of the fermion self-energy can be written as
\begin{equation}\label{eq:paramself}
Re\Sigma'=-{\not K}(a_{L}L+a_{R}R)-{\not u}(b_{L}L+b_{R}R).
\end{equation}
This fermion self-energy, as it was shown in
\cite{Weldon1982,Quimbay1995}, leads to obtain a dispersion
relation for fermions of each chirality by mean of independent
expressions, {\it i. e.} the fermion dispersion relations for
different chirality are not coupled and they are given by
\begin{eqnarray}
\left[ \omega (1 + a_L ) + b_L \right]^2 - k^2 \left[ 1 + a_L \right]^2 = 0,\label{eq:drlh} \\
\left[ \omega (1 + a_R ) + b_R \right]^2 - k^2 \left[ 1 + a_R
\right]^2 =0. \label{eq:drrh}
\end{eqnarray}
As we are interested in the propagation of left-handed neutrinos,
we will only calculate the left-handed invariant functions that
contribute to the left-handed neutrino dispersion relation.\par

We calculate the Lorentz invariant functions by mean of the
following expressions \cite{Weldon1982,Quimbay1995}:
\begin{eqnarray}
a_L=&&\frac1{4\kappa^2}\left[Tr\{{\not K}Re\Sigma'\}-\omega
Tr\{{\not
u}Re\Sigma'\}\right],\label{eq:invarianta}\\
b_R=&&\frac1{4\kappa^2}\left[(\omega^2-\kappa^2)Tr\{{\not
u}Re\Sigma'\}- \omega Tr\{{\not
K}Re\Sigma'\}\right],\label{eq:invariantb}
\end{eqnarray}
where we have defined the Lorentz invariants $\omega$ and $\kappa$
as $\omega=K\cdot u$ and $\kappa=((K\cdot u)^2-K^2)^{1/2}$. These
variables are known in the literature as the invariant energy and
invariant momentum, respectively. With these definitions, the
problem of finding the thermal self-energy is reduced to computing
the traces $Tr\{{\not K}Re\Sigma'\}$ and $Tr\{{\not
u}Re\Sigma'\}$. The calculation of these traces by using the
diagrams of figure \ref{figura1} is relatively long and it
proceeds in a similar way to the calculations shown in
\cite{Quimbay1995,Quimbay1999}. Thus, we restrict ourselves to
show the results for the traces in terms of integrals of the
thermal distribution functions.

The thermal background interacting with left-handed neutrinos is
constituted by massive charged leptons, massive scalar bosons and
massive $W_L$ and $Z_L$ gauge bosons. It is interesting to note
that the initial steps of the calculation readily show that no
flavor change can result from finite temperature effects at least
at one-loop order in perturbation theory. Thus, the incoming and
outgoing neutrino flavors in the diagrams are the same. Therefore,
we can specify each self-energy diagram with three labels: the
external neutrino flavor $i$, the internal lepton $I$, and the
internal gauge boson $B$, with corresponding masses $m^{(0)}_i$
($\sim 0$), $M_I$ and $M_B$, respectively. In the general case, it
is expected for the self-energy to depend on $i$, so we label the
traces with this index. The direct calculation of $Tr\{{\not
K}Re\Sigma'_i\}$ and $Tr\{{\not u}Re\Sigma'_i\}$ in the Feynman
gauge ($\xi=1$) yields:
\begin{eqnarray}
Tr\{{\not K}Re\Sigma_i'\}=&&\frac{8}{(2\pi)^2}\sum_{B,I}C^i_{B,I}
\int_0^\infty d\rho\rho^2\left[\frac{n_b(\epsilon_B)}{\epsilon_B}+
\frac{n_f^-(\epsilon_I)}{2\epsilon_I} +\frac{n_f^+(\epsilon_I)}{2
\epsilon_I}\right]\nonumber\\&&+\frac{2\omega}{(2\pi)^2}\sum_{B,I}
\frac{C'^i_{B,I}}{M^2_{B}}\int_0^\infty
d\rho\rho^2[n^-_f(\epsilon_I)-n^+_f(\epsilon_I)],
\label{eq:tracevecK}\\
Tr\{{\not u}Re\Sigma_i'\}=&&\frac{2}{(2\pi)^2}\frac4{2\kappa}
\ln\left(\frac{\omega+\kappa}{\omega-\kappa}\right)
\sum_{B,I}C^i_{B,I}\int_0^\infty d\rho\rho\left[n_b(\epsilon_B)+
\frac{n_f^-(\epsilon_I)+n_f^+(\epsilon_I)}{2}\right],\label{eq:tracevecu}
\end{eqnarray}
In these traces, the parameter $\epsilon_{I,B}$ is given by
$\epsilon_{I,B}=\sqrt{\rho^2+M_{I,B}^2}$. The coefficients
$C_{B,I}$ and $C'_{B,I}$ come from the respective vertex factors
of the Feynman diagrams. In terms of the gauge group generators,
these coefficients are written as
\begin{eqnarray}
C^i_{B,I}=&&2g^2t^B_{Ii}t^B_{iI},\label{eq:ce}\\
C'^i_{B,I}=&&g^2t^B_{ii}t^B_{II},\label{eq:ceprime}
\end{eqnarray}
Up to this point, we have not sed the explicit constants for the
LRSM. So these expressions can be evaluated straightforward for
any non-abelian gauge theory. We must note that in the unbroken
phase of non-abelian gauge theories with vanishing chemical
potentials, the integrals can be extracted from the diagram sums
and these simplify to quadratic Casimir invariants. The explicit
results for this situation in the LRSM will be shown later. All
the calculations presented here have been performed in the Feynman
gauge, however the case of a generalized $\xi$-gauge will be
discussed briefly later on. It is necessary to point out that the
last term of expression (\ref{eq:tracevecK}) corresponds to the
contribution of the tadpole diagram represented in Fig.
\ref{figura1}. This tadpole diagram is different from zero only
for non-vanishing leptonic chemical potentials and massive gauge
bosons. The contributions coming from the diagrams involving
scalars have the same form that those from gauge diagrams, but in
this case the constants $C$ are given in terms of the
Yukawa-Majorana constants, divided by 2, that couple neutrinos and
scalars, while the constants $C'$ vanish because the scalar
tadpole in Figure \ref{figura1} is identically zero.

\par The analytic evaluation of the integrals appearing in
(\ref{eq:tracevecK}) and (\ref{eq:tracevecu}) is complicated and
generally one needs to resort to numerical techniques or
approximations based on series. For the last case, some
approximations are frequently convergent only in a limited range
of temperatures and/or densities. As a generalization of this kind
of approximations, in the following we will present a method that
is founded on the Mellin transform
\cite{Landsman1987,Flajolet1995} and that permits to obtain series
that converge for any value of the parameters. To outline the
method we will evaluate the integral
\begin{equation}\label{eq:aevaluar}
I_1=\int_0^\infty d\rho \rho^2\frac{n(\rho)}{2\epsilon},
\end{equation}
where $\epsilon=\sqrt{\rho^2+M^2}$ and
\begin{equation}
n(\rho)=\frac1{e^{\beta(\epsilon-\mu)}-\eta}+\frac1{e^{\beta(\epsilon+\mu)}-\eta}.
\end{equation}
The parameter $\eta$ tells us the kind of distribution we are
using. Specifically $\eta=-1$ for the fermion distribution and
$\eta=+1$ for the boson distribution. By using the geometric
series, it is possible to write this integral as
\begin{eqnarray}
I_1=&&\sum_{n=1}^\infty\eta^{n-1}\frac{1}{2\beta^2}\int_{x-y}^\infty
du\sqrt{(u+y)^2-x^2}e^{-nu}+\{y\rightarrow-y\},
\end{eqnarray}
where we have performed the change of variable $x=\beta M$ and
$y=\beta\mu$. The inner integral is evaluated in terms of a
Laplace transformation. Substituting its value and simplifying it,
we find that
\begin{equation}\label{eq:doublesum}
I_1=\sum_{n=1}^\infty\eta^{n-1}\frac{n^{-1}}{\beta^2}x\cosh(ny)K_1(nx),
\end{equation}
where $K_1$ represents the modified Bessel function of order 1.
Essentially, the expression (\ref{eq:doublesum}) corresponds to
the integral that we are looking for. However, this expression is
not convenient to approach conveniently our problem because it
involves a double series which converges slowly. To solve this
problem, we rewrite the series in terms of a simple series by
using the Mellin resummation method \cite{Landsman1987}. This
method is based on the use of the Mellin integral transform
\cite{Flajolet1995,Friot2005}. This method is based on the use of
the following identity
\begin{eqnarray}
\sum_{n=1}^\infty\eta^{n-1}n^{-\nu}f(nx)=&&\int_{c-i\infty}^{c+i\infty}
\frac{ds}{2\pi i}(1-(1-\eta)^{1-s})
\zeta(s)f^*(s-\nu)x^{-s+\nu},\label{eq:identity}
\end{eqnarray}
where $\zeta$ is the Riemann zeta function and the integration is
performed over a line in the complex plane, with $Re[z]=c$. The
function $f^*$ is the Mellin transform of $f$ and is defined as
\begin{equation}\label{eq:Mellin}
f^*(s)=\int_0^\infty f(x)x^{s-1}dx.
\end{equation}
The identity (\ref{eq:identity}) together with
(\ref{eq:doublesum}) allow us to write the integral $I_1$ as an
integral in the complex plane. If certain conditions of asymptotic
behavior are fulfilled, we can consider this complex integral as a
contour integral which is evaluated by using the residue theorem
\cite{Flajolet1995,Friot2005}. This evaluation is in general a sum
over the residues which is the series that we are looking for. The
Mellin transformation of the function $f(x)=\cosh(bx)K_1(x)$, with
$b=\mu/M$ and $K_1$ the modified Bessel function of order 1, can
be written as \cite{Oberhettinger1974}
\begin{eqnarray}
f^*(s)=&&\frac{2^s}{4}\Gamma\left(\frac{s+1}{2}\right)\Gamma\left(\frac{s-1}{2}
\right)\times_2F_1\left(\frac s2-\frac12,\frac s2+\frac12
\frac12;b^2\right),
\end{eqnarray}
where $\Gamma$ is the gamma function and $_2F_1$ is the Gaussian
hypergeometric function. In this work, we limit ourselves to show
the final result for $I_1$. Given that the series depends on the
poles of the integrand in (\ref{eq:identity}), the expansion of
the integral depends strongly on $\eta$. Thus, the integral $I_1$
can be written as
\begin{eqnarray}
I_1=&&\sum_{l=0}^1\sum_{n=\delta_{l,1}}^l(-1)^n(4^{n-l}-(1-\eta)4^{n-l})(2\pi)^{2-2l}
\frac{B_{2-2l}(n-l)!}{(2-2l)!(l-n)!(2n)!}x^{2l-2n}y^{2n} \nonumber\\
&&-(1+\eta)\frac\pi2(x^2-y^2)^{1/2}-\frac\eta2x^2\left[\log\frac{x}{4\pi}+(1-\eta)
\log2+\gamma-\frac12\right]\nonumber\\
&&+2\sum_{l=1}^\infty\sum_{n=0}^l(-1)^{l+1}(4^{n-l-1}-(1-\eta)4^{n-1})
\frac{(2\pi)^{-2l}(2l)!\zeta(2l+1)}{(2n)!(l-n)!(l-n+1)!}x^{2+2l-2n}y^{2n},
\label{eq:fullseries}
\end{eqnarray}
where $B_{2-2l}$ are the Bernoulli numbers. This series is valid
for any value of temperature, masses and leptonic chemical
potential.
\par It is possible to obtain series of the form (\ref{eq:fullseries}) for all the
integrals appearing in (\ref{eq:tracevecK}) and
(\ref{eq:tracevecu}). The convergence of these series is faster at
high temperatures, so the partial sums approximate better the full
expression when $\beta\rightarrow0$. In high temperature limit, we
can approximate (\ref{eq:tracevecK}) and (\ref{eq:tracevecu}) as
\begin{eqnarray}
Tr\{{\not
K}Re\Sigma'\}=&&\frac{4}{(2\pi)^22\beta^2}\sum_{B,I}C_{B,I}\left[\frac{\pi^2}{2}-
\pi x_B+\frac{y_I^2}{2}+\frac{2\gamma-1}{4}(x_I^2-x_B^2)\right]\nonumber\\
&&-\frac{2\omega}{(2\pi)^2\beta}\sum_{B,I}\frac{C'_{B,I}}{x^2_B}\left[\frac{\pi^2}{6}y_I+
\frac{1}{64}y_I\left(\frac{2}{3}y^2_I-x_I^2\right)\right],\\
Tr\{{\not
u}Re\Sigma'\}=&&\frac4{(2\pi)^22\kappa\beta^2}\log\left(\frac{\omega_+}{\omega_-}
\right)\sum_{B,I}C_{B,I}\left[\frac{\pi^2}4-\frac\pi2x_B+\frac{y^2_I}4+\frac{2\gamma-1}{8}
(x_I^2-x_B^2)\right],
\end{eqnarray}
where we have defined $\omega_\pm=\omega\pm\kappa$. From these
expressions and (\ref{eq:invarianta}) and (\ref{eq:invariantb}),
it is found that the Lorentz invariant functions $a$ and $b$ are
given by
\begin{eqnarray}
a=&&-\frac{M'^2}{\kappa^2}\left[\frac{\omega}{2\kappa}\log\left(\frac{\omega_+}{\omega_-}
\right)-1\right],\label{eq:avalue}\\
b=&&-\frac{M'^2}{\kappa^2}\left[\omega-\frac{\omega^2-\kappa^2}{2\kappa}\log\left(
\frac{\omega_+}{\omega_-}\right)\right],\label{eq:bvalue}
\end{eqnarray}
where the constant $M'^2$ is written as
\begin{eqnarray}
M'^2(\beta)=&&\frac{1}{(2\pi)^22\beta^2}\sum_{B,I}C_{B,I}\left[\frac{\pi^2}{2}-\pi\beta
M_B +
\frac{\beta^2\mu_I^2}{2}+\frac{2\gamma-1}4\beta^2(M^2_I-M_B^2)\right]\nonumber\\
&&-\frac{2\omega}{(2\pi)^2\beta^2}\sum_{B,I}\frac{C'_{B,I}}{M^2_B}\left[\frac{\pi^2}{6}\mu_I+
\frac{\beta^2}{64}\mu_I\left(\frac{2}{3}\mu^2_I-M_I^2\right)\right].\label{eq:thermalmass}
\end{eqnarray}
This constant contains all the information about the interactions
that neutrinos have with the medium. However, it can not be
regarded as an effective thermal mass because of the dependency on
the invariant energy $\omega$ induced by the tadpole contribution.
In spite of this, $M'$ reduces to reported results for thermal
effective masses in the limit of vanishing masses and chemical
potentials \cite{Weldon1982,Quimbay1995,Quimbay1999}.

\par It is well known that the dispersion relation can be found by
determining the poles of the propagator corrected with the
fermionic self-energy. For the massless fermions, it has been
shown \cite{Weldon1982} that the dispersion relation for each
chirality is given by the solution of
\begin{equation}\label{eq:basicdispersion}
\omega(1+a_{L,R})+b_{L,R}=\pm\kappa(1+a_{L,R}),
\end{equation}
where the positive and negative signs refer to quasifermions
associated to neutrinos and quasiantifermions associated with the
antineutrinos in the plasma, respectively. In the literature, the
dispersion relations for each case are known as the normal and
abnormal branches. Restricting to the left chirality, it is
possible to obtain an expression that involves both dispersion
relations
\begin{equation}\label{eq:generaldispersion}
\omega\pm\kappa=\frac{M'^2}{2\kappa}\log\left(\frac{\omega_+}{\omega_-}\right)
\left(1\mp\frac\omega\kappa\right)\pm\frac{M'^2}\kappa.
\end{equation}
In this last expression, we have neglected infrared-divergent
terms that arise from the fact that the traces
(\ref{eq:tracevecK}) and (\ref{eq:tracevecu}) are no longer
proportional to each other, as it happens in the massless case. As
the resummation technique has shown that the calculation is
infrared-safe \cite{Kapusta1989}, we ignore those divergent terms
and define the finite part by demanding that the results reduce to
the known expressions for the massless case in such a limit.


\section{Neutrino effective thermal masses\label{sec:effectivemasses}}

\par The quantity that appears in the right hand side of
(\ref{eq:generaldispersion}) can be defined as an effective
potential that acts on the (anti)neutrinos. For the neutrino case
and following a similar procedure as shown in \cite{Weldon1982},
it is possible to find that the dispersion relation in the small
momenta limit ($\kappa \ll \omega$) can be approximated by
\begin{equation}\label{eq:smallmomenta}
\omega=(M^2+m'\omega)\left(1+\frac23\frac\kappa\omega+
\frac{\kappa^2}{\omega^2}+O\left(\frac{\kappa^3}{\omega^3}\right)\right),
\end{equation}
where
\begin{eqnarray}
M^2=&&\frac{1}{(2\pi)^22\beta^2}\sum_{B,I}C_{B,I}\left[\frac{\pi^2}{2}-
\pi\beta M_B
+\frac{\beta^2\mu_I^2}{2}+\frac{2\gamma-1}4\beta^2(M^2_I-M_B^2)
\right],\label{eq:centralmass}\\
m'=&&-\frac{2}{(2\pi)^2\beta^2}\sum_{B,I}\frac{C'_{B,I}}{M^2_B}
\mu_I\left[\frac{\pi^2}{6}+ \frac{\beta^2}{64}\left(\frac{2}{3}
\mu^2_I-M_I^2\right)\right],\label{eq:littlem}
\end{eqnarray}

From the zero momentum limit of (\ref{eq:smallmomenta}), we obtain
that the effective thermal masses for quasifermions $M_{+}$ and
quasiantifermions $M_{-}$ are
\begin{equation}
M_\pm=\frac12(m'\pm\sqrt{m'^2+4M^2}). \label{eq:mgtem}
\end{equation}
We can observe that these effective thermal masses are different
from each other. This fact is reflecting the asymmetry between
number of fermions and antifermions which is described by the
fermionic chemical potential. We remarkably notice that the
quantity $M^2$ provides the effective mass for both particles and
antiparticles in the vanishing chemical potential case, since $m'$
also vanishes.

\par We notice that the expression given by (\ref{eq:mgtem}) is the most
general fermionic effective thermal mass that is possible to
obtain. This effective thermal mass depends on the fermion mass
($M_I$), the gauge boson mass ($M_B$), the leptonic chemical
potential ($\mu_I$) and the temperature ($T=1/\beta$). We observe
that if we fix $M_F=M_B=\mu_I=0$, the expression (\ref{eq:mgtem})
leads us to
\begin{equation}
M_\pm=M=\frac{g^2 C(R)}{8}T^2.
\end{equation}
This result corresponds to the effective mass for the case of a
massless fermion interacting through a massless gauge boson mass,
being $C(R)$ the quadratic Casimir invariant of the representation
defined by $(L^A L^A)_{mn}= C(R)\delta_{mn}$, in agreement with
\cite{Weldon1982}.

Now if we fix $M_F=M_B=0$ into the expression (\ref{eq:mgtem}),
for this case the tadpole contribution does not exist and $m'=0$.
For this reason the expression (\ref{eq:mgtem}) leads us to
\begin{equation}
M_\pm=M=\frac{g^2 C(R)}{8}\left( T^2 +
\frac{\mu_I^2}{\pi^2}\right),
\end{equation}
in agreement with \cite{Quimbay1999}.

\par At this point, it is important to discuss further the validity of
neglecting the term proportional to the neutrino mass in
(\ref{eq:anotherparamself}). For simplicity, we will ignore the
contribution due to the tadpole diagram, in such a way that the
analysis is accurate in the vanishing chemical potential regime.
The properties of the dispersion relation for massive neutrinos
have been discussed in \cite{Petitgirard1992}, therein it was
found that for small momenta the normal and abnormal branches for
neutrinos of mass $m_\nu$ can be written as
\begin{eqnarray}
\omega=&&M'_++\frac1{M^2+M_+^{'2}}\left[\frac{M^2}{3M'_+}+\frac{1}{m_\nu}
\left(\frac{3M_+^2-M^2}{3M'_+}\right)^2\right]K^2+O(K^4),\label{eq:normal}\\
\omega=&&M'_-+\frac1{M^2+M_-^{'2}}\left[\frac{M^2}{3M'_-}+\frac{1}{m_\nu}
\left(\frac{3M_-^2-M^2}{3M'_-}\right)^2\right]K^2+O(K^4),\label{eq:abnormal}
\end{eqnarray}
where $M$ is the effective thermal mass at zero chemical potential
defined as above and
\begin{eqnarray}
M'_+=\frac12\left[(m_\nu^2+4M^2)^{1/2}+m_\nu\right],\label{eq:emplus}\\
M'_-=\frac12\left[(m_\nu^2+4M^2)^{1/2}-m_\nu\right].\label{eq:emminus}
\end{eqnarray}
Considering the present bounds on neutrino masses ($\sum
m_\nu\sim0.1$ eV) and assuming that the temperature of the system
of interest is at least of order $10^9$ K $\sim10^5$ eV, we can
expect safely that $M\gg m_\nu$. Using the binomial series in
expressions (\ref{eq:emplus}) and (\ref{eq:emminus}), we can find
that
\begin{eqnarray}
M'_\pm\approx
M\left(1+\frac18\frac{m_\nu^2}{M^2}\pm\frac{m_\nu}{M}\right).
\end{eqnarray}
Thus, using ($\ref{eq:normal}$) and ($\ref{eq:abnormal}$), we see
that for vanishing momenta the normal and abnormal branches both
tend to the thermal effective mass (\ref{eq:thermalmass}) because
the terms of order $m_\nu$ can be neglected. In this form this
argument validates our approximation. In the following section, we
will apply the formula obtained here for the specific case of LRSM
left-handed neutrinos by expressing the explicit couplings and
diagrams corresponding to each symmetry breaking phase.


\subsection{Effective thermal masses for the unbroken phase}

\par In the unbroken phase all the fermions and gauge bosons
described by LRSM are massless. For the special case in which all
leptonic chemical potentials vanish, the neutrino effective
thermal masses simplify considerably. The explicit form of the
neutrino thermal self-energy depends on the diagrams in which the
neutrino is involved. The expressions (\ref{eq:avalue}) and
(\ref{eq:bvalue}) specify the final form of the effective thermal
masses for the left-handed neutrinos. The diagrams with an
exchange of charged scalar and $W_L^{\pm}$ charged electroweak
gauge bosons induce a flavor change in the incoming neutrino $I$
to a different outgoing neutrino $F$. In the latter contributions,
the flavor $i$ of the internal charged lepton (inside the loop)
runs over the three lepton flavors. The one-loop contribution to
the real part of the self-energy lead us to write the squared
value of the effective thermal masses as
\begin{eqnarray}
M^{2}_{IF}=\frac{T^2}{8}\left[\left(\frac32g^2+g'^2\right)\delta_{IF}+
\frac{3}{2} \left[(ff^t)_i+(\tilde h\tilde h^t)_i +(\tilde g\tilde
g^t)_i\right]_{IF}\right].\label{eq:unbrokenmass}
\end{eqnarray}
The different Yukawa-Majorana coupling constants for the distinct
neutrino flavors introduce the flavor non-degeneracy of the
branches that describe the quasi-particles. The neutrino effective
masses are non-degenerate since the Yukawa-Majorana coupling
constants are different for the different flavors. The differences
between the neutrino effective masses can not be unworthy and they
can affect the left-handed neutrino oscillations.

\par In order to get a deeper insight about these results in the LRSM
context, it is appropriate to compare the thermal effective masses
given by (\ref{eq:unbrokenmass}) with their counterpart of the ESM
without chemical potentials. After some simplification, we obtain
that the neutrino effective masses in the ESM are given by
\begin{equation}
M^2_{i}=\frac{T^2}{8}\left(\frac34g^2+g^{'2}+|f|^2_i\right).
\end{equation}
Thus, it is possible to reproduce the results of \cite{Weldon1982}
but with an effective thermal mass for each neutrino species.
These effective masses can be written in terms of the charged
lepton masses $m_i$ and the charged electroweak gauge boson mass
$M_{W_L}$ as
\begin{eqnarray}
M^{2}_i=&&\frac{T^2}{8}\frac{g^2}{4}\left(2+
\frac{1}{c^2_w}+\frac{m^2_i}{M^2_{W_L}}\right).\label{eq:unbrokenmass1}
\end{eqnarray}
in agreement to a previous computation performed in
\cite{Quimbay1995}. In this reference, the neutrino dispersion
relations for the electroweak unbroken phase of the ESM were
calculated explicitly.


\subsection{Effective thermal masses for the parity-broken phase}

\par In the phase where the temperature is low enough so that the
original gauge symmetry is broken down to the electroweak group
$SU(2)_L\times U(1)_Y$, the left-handed neutrinos do not possess
mass but the contributions to the thermal effective mass that
involve the right-handed particles begin to be thermally
suppressed. Considering the case of the integral
(\ref{eq:doublesum}), we see that in the limit $x=\beta
M\rightarrow\infty$ the modified Bessel function $K_1(nx)$ behaves
asymptotically as $\exp{(-nx)}$. Thus the contribution from this
integral to the self-energy tends to zero. The same property is
satisfied by the remaining integrals shown in (\ref{eq:tracevecK})
and (\ref{eq:tracevecu}). Therefore, it can be concluded that at
temperatures $T\ll v_R$ we can neglect the $T\neq0$ contribution
of the diagrams involving the right-handed particles. By this
reason, the calculation of the thermal effective masses for this
case is similar to the one performed in the unbroken electroweak
symmetry case, but using the couplings for the left-handed
particles given in the Lagrangian densities
(\ref{eq:gaugeleftright}) and (\ref{eq:yukawaleftright}).

\par When the left-handed symmetry is unbroken, the
left-handed gauge bosons do not have mass. Furthermore, because
none of the left-handed fermions have mass then the contribution
of the gauge diagrams involving these particles is nearly the same
as in the unbroken electroweak case, but including fermionic
chemical potentials. Thus the thermal effective mass of the
$i$-flavor neutrino can be written as
\begin{equation}
M^{(g)2}_{i}=\frac{T^2}{8}\left(\frac32g^2+g^{'2}\right)+
\frac{1}{8\pi^2}\left[g^2(\mu_e^2+\mu_\mu^2+\mu_\tau^2)+
\frac12g^2(c_w+s_wt_w)^2\mu^2_i\right],
\end{equation}
where $\mu_i$ is the fermionic chemical potential associated to
the $i$-flavor neutrino. To obtain this thermal effective mass, we
have also gotten that the tadpole contribution mediated by $Z$
vanishes. The only non-vanishing tadpole term involves the $Z'$
boson, but this term is negligible in the regime in which we are
working at ($T\ll v_R$). Hence, for the regime $m'\approx 0$, we
obtain that quasiparticles and quasiantiparticles have nearly the
same effective mass.

\par With respect to the scalar contribution, the structure of the
effective thermal mass is heavily influenced by the pattern in
which the different scalar fields gain masses and mixings. By
counting the degrees of freedom of the scalar multiplets
introduced to break the gauge symmetry of the LRSM, we obtain that
there are 20 scalar fields involved in this dynamics. In order to
give mass to scalar fields, a scalar potential is introduced and
includes all possible terms respecting the gauge and Lorentz
invariance. The complete form of this potential is complicated
\cite{Barenboim2001}, so we will not enter into details here.
Given the size of the matrices involved the analytic computation
of scalar masses and mixings is very difficult in general. We will
use the results of a numerical analysis presented in
\cite{RodriguezY2002}, so we can get an expression for effective
thermal masses in this context.

\par It has been shown in \cite{RodriguezY2002} that there are
flavour changing neutral currents (FCNC) in the LRSM associated
with the charged scalar bosons. One possible way to avoid these
FCNC, without performing any fine tuning on the coupling constants
or the vacuum expectation values, is to give a really heavy mass
to the scalar bosons \cite{RodriguezY2002}. Then, we can consider
a model with FCNC suppressed enough to be consistent with the
experimental constraints, and with the additional feature of
having a spontaneous origin for the observed CP violation. The
scalar spectrum depends on the CP-violating phases. For instance,
with $\alpha=0$ and $\theta=\pi/2$, we have 3 neutral bosons with
masses of order $v_R$ while the remaining neutrals are massless
\cite{RodriguezY2002}. For the situation in which both phases
vanish, the number of massive neutral scalars increases to 5. In
this work, we limit ourselves to work with $\alpha=0$ and
$\theta=\pi/2$ scenario, in which there are more Higgs bosons at
low energies than in the ESM. With this, and using the values for
scalar masses and mixing given by tables 7, 8 and 9 of
\cite{RodriguezY2002}, we can write the scalar contribution to the
effective thermal masses as
\begin{eqnarray}
M^{(s)2}_i=&&\frac{T^2}{16}\left(3.00(\tilde h\tilde
h^\dagger)_{ii}+
(\tilde g\tilde g^\dagger)_{ii}+4.14(ff^\dagger)_{ii}\right)\nonumber\\
&&+\frac12\sum_j\left[\tilde h_{ij}\tilde
h_{ji}^*(2.00\mu_{\nu_j}^2+ \mu_{l_j}^2)+\tilde g_{ij}\tilde
g_{ji}^*\mu_{l_j}^2+f_{ij}
f_{ji}^*(1.23\mu_{\nu_j}^2+3.09\mu_{l_j}^2)\right].
\end{eqnarray}
We see that the additional scalar bosons have a decisive effect in
the thermal mass. In contrast with the former case, we have that
this result can be compared with the ESM predictions for contexts
where the electroweak symmetry is restored as, for example, in the
early universe.


\subsection{Effective thermal masses in the fully-broken phase}

Here we will present the thermal effective masses for the
completely broken phase of the LRSM. It is remarkable that the
formula, which we have presented before for the case of massive
particles, allow us to obtain fully analytic results for this
regime. Direct evaluation of equations (\ref{eq:centralmass}) and
(\ref{eq:littlem}), with the diagrams involving gauge bosons and
neutrino species $i$, yield to
\begin{eqnarray}
M^{(g)2}_I=&&\frac{T^2}{8}\left(\frac32g^2+g'^{2}\right)+\frac{1}{2\pi^2}
\left[-\pi
T\left(\frac{g^2}{2}M_W+\frac{g^2}{4c_w^2}M_Z\right)+\frac12
\left(\frac{g^2}{2}\sum_jU_{ij}U_{ji}^*\mu_{l_j}^2\right.\right.\nonumber\\
&&\left.\left.+\frac{g^2}{4c_w^2}\sum_jU_{ij}U_{ji}^*\mu_{\nu_j}^2\right)+
\frac{2\gamma-1}{4}\left(\frac{g^2}{2}\sum_jU_{ij}U_{ji}^*M_{l_j}^2
+\frac{g^2}{4c_w^2}\sum_jU_{ij}U_{ji}^*M_{\nu_j}^2\right.\right.\nonumber\\
&&\left.\left.-\frac{g^2}{2}M^2_W-\frac{g^2}{4c_w^2}M^2_Z\right)\right],
\end{eqnarray}
where we have taken into account the possibility of neutrino
mixing through the MNS mixing matrix $U$. For the case of the ESM,
it is relevant that this constant has the same form up to some
coupling constants given that the particle spectrum with which we
are working is the same. The $Z$-mediated tadpole diagram
contribution takes the form
\begin{eqnarray}
m'_i=&&-\frac{T^2g^2}{2\pi^2M_Z^2}\left[\frac{\pi^2}{6}\sum_j
\left(\mu^2_{l_j}+\frac12(c_w+s_wt_w)^2\mu_{\nu_j}\right)\right.
\nonumber\\
&&\left.+\frac{1}{64T^2}\sum_j\left(\mu_{l_j}\left(\frac23
\mu_{l_j}^2-M_{l_j}^2\right)+\frac26(c_w+s_wt_w)^2\mu_{\nu_j}^3\right)\right],
\end{eqnarray}
where $t_w=\tan(\theta_w)$. It is interesting to note that this
quantity takes the same value for all neutrino species. As in the
parity-broken case, we expect that the main differences with the
ESM arise from the larger Yukawa sector of the LRSM. Restricting
ourselves to the same regime as the preceding section, we can
write the scalar contribution as
\begin{eqnarray}
M^{(s)2}_i=&&\frac{T^2}{2(2\pi)^2}\left[\frac{\pi^2}{2}
\left(3.00(\tilde h\tilde h^\dagger)_{ii}+(\tilde g\tilde
g^\dagger)_{ii}
+4.14(ff^\dagger)_{ii}\right)\right.\nonumber\\
&&-\frac\pi T\left((\tilde h\tilde
h^\dagger)_{ii}(0.50M_{\phi_D^0}+M_{\phi_F^0})+
(ff^\dagger)_{ii}(0.25M_{\delta_L^+}
+M_{\delta^{++}_L}+0.50M_{\phi_D^0}+M_{\phi_E^0})\right)\nonumber\\
&&+\frac{1}{2T^2}\sum_j\left(\tilde h_{ij}\tilde
h^*_{ji}(2.00\mu^2_{\nu_j}+\mu^2_{l_j})+ \tilde g_{ij}\tilde
g^*_{ji}\mu^2_{l_j}+
f_{ij}f_{ji}^*(1.23\mu^2_{l_j}+3.09\mu^2_{l_j})\right)\nonumber\\
&&+\frac{2\gamma-1}{4T^2}\left(\sum_j(\tilde h_{ij}\tilde
h_{ji}^*+\tilde g_{ij} \tilde
g_{ji}^*+3.09f_{ij}f_{ji}^*)M_{l_j}^2
+\sum_j(2\tilde h_{ij}\tilde h^*_{ji}+1.24f_{ij}f_{ji}^*)M_{\nu_j}^2\right.\nonumber\\
&&\left.\left.-(\tilde h\tilde h^\dagger)_{ii}(0.5M^2_{\phi_D^0}+
M^2_{\phi_F^0})-(ff^\dagger)(0.25M_{\delta_L^+}^2
+M_{\delta^{++}_L}^2+0.50M_{\phi_D^0}^2+M_{\phi_E^0}^2)\right)\right],
\end{eqnarray}
where we have used the nomenclature used in \cite{RodriguezY2002}
for the scalar fields with definite mass. We note that these
results are accurate in the case in which the temperatures are
bigger than the masses of the particles. Even if the general
series obtained before are convergent for any value of
temperature, in the low $T$ regime the contribution of the
imaginary part of the self energy must be taken into account. For
this case (\ref{eq:tracevecK}) and (\ref{eq:tracevecu}) are not
enough to describe the propagation of quasiparticles.

\par All these results have been calculated at Feynman gauge, however
it is necessary to know if the results hold in other gauges. In
the case of unbroken symmetry, former studies
\cite{Quimbay1995,Quimbay1999} indicate that the self-energy
calculated at one loop order is gauge invariant if just leading
terms in temperature are kept. For the general case of broken
symmetry, this property generally does not hold. But it has been
shown \cite{DOlivo1992} that although the neutrino self-energy can
be gauge-dependent. In that case, the dispersion relation only
receives gauge-dependent contributions at higher order on the
coupling constants. So for the approximation regime that we have
implemented here, we can state that the dispersion relation
(\ref{eq:generaldispersion}) is gauge invariant.


\section{Conclusions\label{sec:conclusions}}

\par We have calculated the left-handed neutrino dispersion relations at
finite temperature and density on the framework of the LRSM
considering that there exists an excess of leptons over
antileptons in the medium. In this context, we have obtained the
neutrino effective thermal masses from the unbroken, parity-broken
and fully-broken phases. These mentioned results were possible to
be obtained due to our first calculations for the most general
neutrino thermal self-energy using the Mellin summation technique.
This last calculation has been performed using the real-time
formalism of thermal field theory at finite temperature and
density. For the first time, we have found a general expression
for the neutrino effective mass which depends on fermion masses,
gauge boson masses, leptonic chemical potential and temperature.
Different effective thermal masses known in the literature can be
obtained as specific cases from this general effective thermal
mass.

\begin{acknowledgements}
The authors would like to thank COLCIENCIAS for financial support
through the project "Propagaci\'on de neutrinos a temperatura y
densidad finita", approved at "Convocatoria J\'ovenes
investigadores e Innovadores - Year 2008". This work also has been
supported by "Divisi\'on de Investigaci\'on of Universidad
Nacional de Colombia", Sede Bogot\'a, by means of the project
"Propagaci\'on y oscilaciones de neutrinos a temperatura y
densidad finita". C. J. Quimbay thanks also Vicerrectoria de
Investigaciones of Universidad Nacional de Colombia by the
financial support received through the research grant "Teoría de
Campos Cuánticos aplicada a sistemas de la Física de Partículas,
de la Física de la Materia Condensada y a la descripción de
propiedades del grafeno".
\end{acknowledgements}

\end{document}